# Decoding Grain Boundary Thermodynamics in High-Entropy Alloys in a 5D Space: Coupled Segregation and Disordering

Chongze Hu[1, 2, #] and Jian Luo[1, 2, *]

[1]*Department of Nanoengineering;* [2]*Program of Materials Science and Engineering,*
*University of California San Diego,*
*La Jolla, California 92093, USA*

*Corresponding author. E-mail: jluo@alum.mit.edu (J.L.)

#Current address: Sandia National Laboratories.

## Abstract

Grain boundaries (GBs) can critically influence the microstructural evolution and various materials properties. However, a fundamental understanding of GBs in high-entropy alloys (HEAs) is lacking because of the complex couplings of the segregations of multiple elements and interfacial disordering, which can generate new phenomena and challenge the classical theories. Here, by combining large-scale atomistic simulations and machine learning, we demonstrate the feasibility of predicting the GB properties as functions of four independent compositional degrees of freedoms and temperature in a 5D space. Subsequently, GB counterparts to bulk phase diagrams are constructed for the first time for quinary HEAs. A data-driven discovery further reveals new coupled segregation and disordering effects in HEAs. Notably, an analysis of a large dataset discovers a critical compensation temperature at which the segregations of all elements virtually vanish simultaneously. While the machine learning model can predict GB properties via a black-box approach, a surrogate data-based analytical model (DBAM) is constructed to provide more physics insights and better transferability, with good accuracies. This study not only provides a new paradigm enabling prediction of GB properties in a 5D space, but also uncovers new GB segregation phenomena in HEAs beyond the classical GB segregation models.

## Graphical Abstract

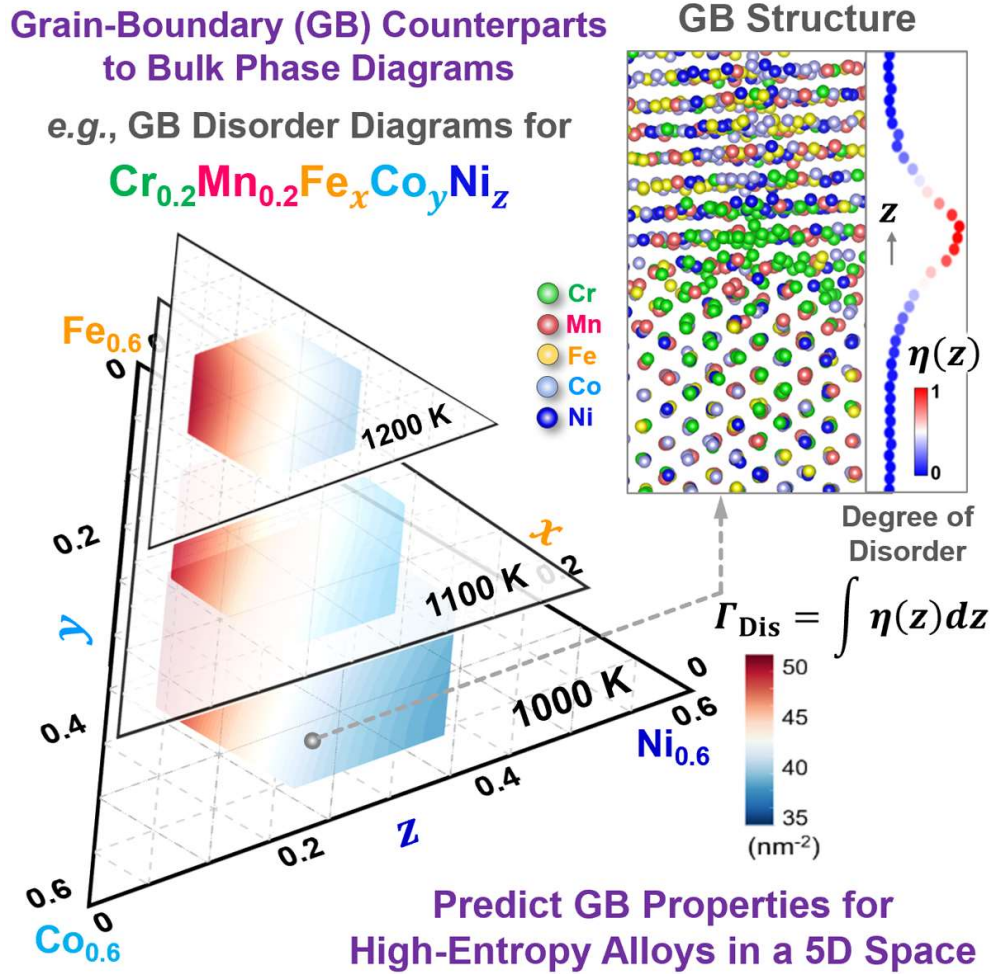

## Progress and Potential

The date-driven prediction of grain boundary (GB) properties in high-entropy alloys (HEAs) as functions of temperature and four independent compositional degrees of freedom in a 5D space opens a new paradigm. The interaction of multiple elements and interfacial disordering can induce a new region of segregation not predicted by the classical theory. A significant data-driven discovery is the existence of a critical compensation temperature with the simultaneous vanishing of GB segregation. A surrogate data-based analytical model can predict GB properties based on parameters with clear physical meanings. This study expands our fundamental knowledge of GB segregation and paves the way for tailoring properties of HEAs via controlling GBs.

**Introduction**

Since the Bronze and Steel Ages, the development of every major class of metallic alloys, such as the Cu, Fe, Al, Ti, and Ni-based alloys, have revolutionized technologies and changed our daily lives. High-entropy alloys (HEAs), also known as multi-principal element alloys (MPEAs) or complex concentrated alloys (CCA), represent the newest class of alloys that attract significant research interest.[1-4] The vast composition space of HEAs offers immense opportunities for designing materials for various applications.

In every class of polycrystalline alloys, grain boundaries (GBs) exist ubiquitously.[5,6] The elemental segregation (*a.k.a.* adsorption in the interfacial thermodynamics) at GBs is a critical phenomenon that can change microstructural evolution[7-9] and govern a broad range of materials properties[10-12]. Even though the GB segregation have been extensively researched for decades, most prior studies and models are based on alloys with one primary (principal) element.[13-20] Moreover, the effects of interfacial disordering on segregation are typically not considered in the classical site-occupying models [13-15]. The GB segregation in the emerging HEAs containing five or more principal elements are hitherto only investigated by few experimental[21,22] and theoretical[23-25] studies (and only for a few compositions). The underlying mechanisms of GB segregation in HEAs are elusive and a predictive model does not exist, which motivate this study.

In a broader perspective, GBs can be considered as two-dimensional (2D) interfacial phases [6], which are also named as "complexions"[26,27] to differentiate them from thin precipitated layers of 3D bulk phases. Notably, GB diagrams, which represent GB thermodynamic states or properties as functions of thermodynamic variables such as temperature and bulk composition (representing chemical potentials), have been developed as the GB counterparts to bulk phase diagrams. To date, various GB diagrams have been constructed for binary and ternary systems,[19,20,26,28] but they are rarely developed for multicomponent systems,[29] certainly not for HEAs, owing to the increasing complexity of a large, multi-dimensional compositional space. Furthermore, the more general GBs (asymmetric GBs with mixed twist and tilt features), which are ubiquitous in polycrystalline materials and often the weak links chemically and mechanically,[28,30] are still scarcely studied.

Herein, by combining the large-scale hybrid Monte Carlo and molecular dynamics (MC/MD) simulations and artificial neutral network (ANN), we demonstrate the feasibility of predicting the GB properties as functions of four independent compositional DOFs and temperature in a 5D space for a representative general GB in $Cr_xMn_yFe_zCo_lNi_m$ HEAs.

Our MC/MD simulations further reveal the unrecognized importance of interfacial disorder in influencing GB segregation that can produce new phenomena in HEAs. Notably, an analysis of the large dataset uncovers the almost (albeit not complete) vanishing of GB segregation/depletion of all elements simultaneously at a critical compensation temperature of ~1388 K for the $Cr_xMn_yFe_zCo_lNi_m$ HEAs.

Furthermore, a data-based analytical model (DBAM) is created (grounded on the analysis of the large dataset generated by hybrid MC/MD simulations) to represent GB segregation and disordering in HEAs analytically. While the machine learning model can make predictions via a black-box approach, this DBAM describes the GB properties in a 5D space via analytical formulae where all parameters have clear physical meanings, and it offers better transferability.

**The Workflow**

Figure 1 displays the workflow of predicting GB properties and GB diagrams, investigating the underlying mechanisms and uncovering new interfacial phenomena, and developing a surrogate DBAM via combined large-scale hybrid MC/MD simulations, ANN, various data analysis approaches, and density functional theory (DFT) calculations. First, we selected an asymmetric Σ81 GB of mixed tilt and twist characters to represent general GBs. We randomly generated 258 compositions out of 1371 possible choices by varying the amount of each element from 5% to 35% with a step of 5%. Second, principal components analysis (PCA) was used to ensure the 258 selected compositions are sufficiently randomly (Supplementary Discussion 1). Third, the large-scale isothermal-isobaric (constant $NPT$) ensemble hybrid MC/MD simulations were carried out to calculate the adsorption amounts (*i.e.*, GB excesses of solutes: $\Gamma_{Cr}$, $\Gamma_{Mn}$, $\Gamma_{Fe}$, $\Gamma_{Co}$, $\Gamma_{Ni}$), GB excess of disorder ($\Gamma_{Dis}$), and GB free volume ($V_{Free}$), as well as bulk composition of each element (to map out a relation between bulk composition and chemical potential), from 1000 K to 1300 K with a temperature step of 100 K. Fourth, the MC/MD-simulated dataset was used to develop an ANN model to predict six GB properties (excluding $V_{Free}$ due to a weak correlation to be discussed later), each in a 5D space. Fifth, GB diagrams of thermodynamic properties were constructed for the first time for quinary HEAs; as an illustrative example, Figure 1E shows isothermal sections of $\Gamma_{Cr}$ for $Cr_xMn_yFe_zCo_{0.2}Ni_{0.2}$ subsystem, where $x + y + z = 0.6$, from 1000 to 1200 K. Sixth, a large dataset generated by hybrid MC/MD simulations was also used to analyze and investigate the new coupled interfacial disordering and segregation phenomena in HEAs beyond the classical models via a data-driven discovery approach. Seventh, additional hybrid MC/MD simulations were conducted for other GBs to show the generality of our findings. Eighth, DFT calculations were conducted, based on the GB configurations obtained by hybrid MC/MD simulations, to investigate the unique segregation mechanisms in HEAs. Ninth, a surrogate DBAM model was developed to predict the GB segregation and disordering to provide more physics insights with better transferability than the "black-box" ANN model, which can also achieve good accuracies. See Computational Procedures for further details.

It is worth noting that before we conducted large-scale hybrid MC/MD simulations, we first performed careful benchmark simulations to validate our $NPT$-based hybrid MC/MD method by comparing it with prior $NVT$-based MC simulations[24] as well as experiments[21,22,31]; see Supplementary Discussion 2.

**ANN prediction of GB diagrams of thermodynamic properties**

The dataset generated by 1032 MC/MD simulations have been used to train, evaluate, and test one-layer single-task ANN models to predict six GB properties ($\Gamma_{Cr}$, $\Gamma_{Mn}$, $\Gamma_{Fe}$, $\Gamma_{Co}$, $\Gamma_{Ni}$, and $\Gamma_{Dis}$).

The histogram of root-mean-square errors (RMSEs) was used to assess the ANN performance. Notably, the ANN models are fairly accurate to predict the values of $\Gamma_{Co}$, $\Gamma_{Fe}$, $\Gamma_{Cr}$, and $\Gamma_{Mn}$ with small RMSEs (Suppl. Figure S4). The accuracies are further supported by the parity plots between ANN predictions and MC/MD simulations, where the promising linear relations are achieved for $\Gamma_{Cr}$, $\Gamma_{Mn}$, $\Gamma_{Fe}$, and $\Gamma_{Co}$ (see Figure 2A and Suppl. Figure S3). Relatively large deviations are found for $\Gamma_{Ni}$ and $\Gamma_{Dis}$, which can be ascribed to the weak segregation of Ni and large uncertainty in quantifying $\Gamma_{Dis}$. Overall, the ANN models are robust to predict GB properties, especially for moderate and strong segregation (*e.g.*, Cr and Mn) and depletion elements (*e.g.*, Fe and Co) at HEA GBs.

To further validate our ANN models, we adopt a structural similarity index (SSIM; 1 = same and 0 = different) to compare the similarity of ANN-predicted binary GB diagrams *vs.* MC/MD simulations; representative GB diagrams are shown in Figure 2B-C. The SSIM histogram (Suppl. Figure S5) shows the high values (~0.88-0.89) for most $\Gamma_{Cr}$, $\Gamma_{Mn}$, $\Gamma_{Fe}$, and $\Gamma_{Co}$ diagrams, but relatively low values (~0.63-0.66) for $\Gamma_{Ni}$ and $\Gamma_{Dis}$ diagrams. This is consistent with the prior analysis based on RMSEs. See further discussion in Supplementary Discussion 3.

Interestingly, ANN predictions can outperform MC/MD simulations in two aspects. First, ANN models can suppress the MC/MD errors caused by the large thermal noises at high temperatures (Suppl. Figure S6) by a smoothing effect. Second, the ANN models become more convenient than MC/MD simulations to predict GB diagrams with multiple variables. For example, Figures 2D-I show the ANN-predicted ternary GB adsorption and disorder diagrams in Cr$_x$Mn$_{0.2}$Fe$_y$Co$_{0.2}$Ni$_z$ (where $x + y + z = 0.6$) at 1000 K. More ANN-predicted GB diagrams can be found in Suppl. Figures S23-S29.

Notably, the efficient ANN models make it possible to map out GB properties as functions of four independent compositional DOFs and temperature in a 5D space for HEAs.

**New GB segregation phenomena in HEAs: A starting example**

Beyond the ANN model prediction, we further conducted a series of in-depth data analyses, along with additional focused simulations, to elucidate new interfacial phenomena that are unique to HEAs. As the first example to illustrate new GB segregation phenomena in HEAs, we have conducted and analyzed MC/MD simulations of nine representative equimolar ternary (medium-entropy) to quinary (high-entropy) alloys, including: FeCoNi, CrMnNi, CrMnFe, CrFeNi, CrCoFe, CrFeCoNi, CrMnFeNi, CrMnFeCo, and CrMnFeCoNi.

As one interesting discovery, we find that the site competition in ordered GBs can lead to weak segregation in ternary alloys, while segregation of multiple elements coupled with GB disordering can induce strong co-segregation in quinary alloys (Figure 3A-B). Specifically, the MC/MD-simulated GB structure of the CrMnNi ternary alloy at 1000 K (Figure 3C) shows that the relatively ordered GB ($\Gamma_{Dis}$ of ~39 nm$^{-2}$) has weak segregation of Cr and virtually no segregation of Mn ($\Gamma_{Cr}$ = 5.3 nm$^{-2}$ and $\Gamma_{Mn}$ = 0.8 nm$^{-2}$). However, the more disordered GB ($\Gamma_{Dis}$ of ~43 nm$^{-2}$) in CrMnFeCoNi exhibits strong co-segregation of Cr and Mn ($\Gamma_{Cr}$ = 18.6 nm$^{-2}$ and $\Gamma_{Mn}$ = 7.0 nm$^{-2}$). The compositional profiles shown in Figure 3C-D confirm the strong GB segregation of both Cr and Mn in CrMnFeCoNi, but weak GB segregation in CrMnNi. Moreover, the computed profile of the disorder parameter also verifies a more disordered GB core of ~0.88 nm thick in CrMnFeCoNi *vs.* a less disordered GB core of ~0.75 nm thick in CrMnNi.

It is interesting to further note that for the systems without Mn element (*e.g.*, CrCoFe and CrFeCoNi), Cr atoms are not favorable to segregate at relatively ordered GBs (Suppl. Figure S8). The structural analysis based on polyhedral template matching (PTM) approach[32] shows that Mn segregation can induce significant GB disordering (Suppl. Figure S8). Since disordered GB can prompt segregation, our observation suggests that the Cr segregation is enhanced by Mn-segregation induced GB disordering. See Supplementary Discussion 4 for more results and further discussion of this set of nine representative equimolar alloys. Notably, all abovementioned phenomena are also found in other ternary, quaternary, and quinary systems, thereby suggesting the generality of the coupling effects.

**Analysis of the large dataset of GB thermodynamic properties**

To have more in-depth understanding of couplings among GB properties, we further calculated Pearson correlation coefficients (PCCs) among five GB adsorption (*i.e.*, $\Gamma_{Cr}$, $\Gamma_{Mn}$, $\Gamma_{Fe}$, $\Gamma_{Co}$, and $\Gamma_{Ni}$) and two structural properties ($\Gamma_{Dis}$ and $V_{Free}$) based on the MC/MD-simulated dataset. Notably, the heat map of PCC shows that GB disorder is strongly correlated with GB adsorption properties. However, there is almost no correlation between GB free volume ($V_{Free}$) with others. Specifically, the segregation of Mn ($\Gamma_{Mn}$) has the strongest correlation with GB disorder ($\Gamma_{Dis}$) among all elements, which agree with the MC/MD simulations showing that Mn segregation can induce large GB disordering. In addition, by calculating the PCCs at different temperatures, we found that the correlations between GB disorder and adsorption properties decrease with increasing temperature, while the correlations between other GB properties remain almost unchanged. The correlation analysis (Supplementary Discussion 5) further verifies the importance of interfacial disordering on GB segregation in HEAs.

The correlations of GB segregations of different elements in HEAs also show interesting trends and suggest new interfacial phenomena. On one hand, GB disordering can promote the segregations of both Cr and Mn (*i.e.*, both $\Gamma_{Cr}$ and $\Gamma_{Mn}$ are positively correlated with $\Gamma_{Dis}$, as shown in Figure 4A); consequently, $\Gamma_{Cr}$ and $\Gamma_{Mn}$ are positively correlated (Figure 4A), despite that a positive Cr-Mn mixing enthalpy (Suppl. Table S3) suggests repulsion between them. On the other hands, a significant negative Co-Cr mixing enthalpy (Suppl. Table S3) suggests that they should attract one another in the bulk phase, but their GB adsorptions are negatively correlated (Figure 4A) because $\Gamma_{Cr}$ is positively, but $\Gamma_{Co}$ is negatively, correlated with $\Gamma_{Dis}$ (Figure 4A). These findings again suggest the critical role of interfacial disordering in influencing GB segregation in HEAs.

Next, we examine the correlation of $\Gamma_i$ vs. $\Gamma_{Dis}$ at different temperatures. Linear regression analyses (Figure 4C and Suppl. Figure S10) show the following statistical relation:

$$\Gamma_i(T,X) - \Gamma_i^0 = \bar{\alpha}_{Dis}^i(T) \cdot [\Gamma_{Dis}(T,X) - \Gamma_{Dis}^0], \qquad (1)$$

where $T$ is temperature, $X = \{X_i\}$ is the bulk composition, $(\Gamma_i^0, \Gamma_{Dis}^0)$ is the intersection point of all linear regression lines that is virtually independent of temperature, and $\bar{\alpha}_{Dis}^i(T)$ is the slope. Interestingly, excellent linear correlations exist for $\bar{\alpha}_{Dis}^i$ vs. $T$ for all elements (Figure 4D):

$$\bar{\alpha}_{Dis}^i(T) = \beta_i \cdot (T - T_C), \qquad (2)$$

where $\beta_i$ is the slope. Notably, the linear regression lines of $\bar{\alpha}_{Dis}^i$ vs. $T$ for all five elements cross over at nearly the same point on the $T$ axis (Figure 4D), which is denoted as $T_C$ (~ 1388 ± 51 K).

Taking Cr as one example, the MC/MD-simulated $\Gamma_{Cr}$ has linear relation with $\Gamma_{Dis}$ statistically (Figure 4C). The positive slopes ($\bar{\alpha}_{Dis}^{Cr} > 0$) of the $\Gamma_{Cr}$ vs. $\Gamma_{Dis}$ regression lines are related to the positive Cr segregation at the GB. The fitted $\bar{\alpha}_{Dis}^{Cr}$ value linearly decays by increasing the temperature with a negative slope of $\beta_{Cr}$, and intersects with the $T$ axis at $T_C = 1347$ K (Figure 4D). Similar behavior can also be found for Mn (with moderate positive GB segregation), where $\bar{\alpha}_{Dis}^{Mn} > 0$ (Suppl. Figure S10B), $\beta_{Mn} < 0$, and $T_C = 1464$ K (albeit a high uncertainty in $T_C$ due to the small slope). In contrast, the slopes of $\Gamma_{Fe(Co)}$ vs. $\Gamma_{Dis}$ regression lines are negative ($\bar{\alpha}_{Dis}^{Fe(Co)} < 0$) due to the depletion of Fe or Co (Suppl. Figure S10C-D); consequently, $\bar{\alpha}_{Dis}^{Fe(Co)}$ linearly increases with increasing temperature ($\beta_{Fe(Co)} > 0$; $T_C = 1370$ K for Fe and 1371 K for Co, respectively, in Figure 4D). Finally, there is only a weak correlation in $\Gamma_{Ni}$ vs. $\Gamma_{Dis}$ with a small negative slope due the small $\Gamma_{Ni}$ values, which also results in (large relative noises in Suppl. Figure S10E) and large error on the projected $T_C$ (due to small slope and large uncertainty).

**A data-based analytical model (DBAM)**

Next, we propose a DBAM as a surrogate model based on above analysis of the large MC/MD-simulated dataset, particularly the linear correlations represented by equations (1) and (2) and shown in Figures 4C-D and Suppl. Figure S10. The detailed derivation and data-fitting of this DBAM can be found in Computational Procedures and Supplementary Discussion 6. In this DBAM, the GB excess of component $i$ ($\Gamma_i$) and disorder ($\Gamma_{Dis}$) as functions of temperature ($T$) and bulk composition ($X = \{X_i\}$) of an HEA can be respectively expressed as:

$$\Gamma_i(T,X) = \beta_i \cdot (T - T_C) \cdot [\Gamma_{Dis}(T,X) - \Gamma_{Dis}^0] + \sum_j(\kappa_{i,j}^{Seg} \cdot X_j), \qquad (3)$$

and

$$\Gamma_{Dis}(T,X) = \sum_i(\kappa_i^{Dis} \cdot X_i) \cdot \exp\left(-\frac{E_A^{Dis}}{k_B T}\right). \qquad (4)$$

Here, $\kappa_{i,j}^{Seg}$ is the coupling coefficient for GB segregation between component $i$ and $j$, $E_A^{Dis}$ is an activation energy, $k_B$ is the Boltzmann constant, and $\kappa_i^{Dis}$ is the coupling coefficient for GB disorder and component $i$. See Computational Procedures and Supplementary Discussion 6 for details. By using the best fitted parameters listed in Tables S1 and S2, the predicted GB properties (all five $\Gamma_i$ and $\Gamma_{Dis}$) from this DBAM agree with MC/MD simulations with a small root-mean square error (RMSE) of ~2.3 nm$^{-2}$ (Suppl. Figure S11).

Here, a distinct merit of this DBAM (in contrast to the ANN model) is that all the model parameters have clear physical meanings. The fitted $\kappa_i^{Dis}$ and $\kappa_{i,j}^{Seg}$ values listed in Suppl. Tables S1 and S2 represent the couplings between segregation and disorder, as well as segregation of different elements, which are fully consistent with the trends observed in our MC/MD simulations, as discussed in Supplementary Discussion 6. Notably, we can predict the $\Gamma_i(T, X)$ with relatively small RMSEs (albeit slightly larger than those from the ANN predictions) for each element using this simple analytical equation (3), as shown in Suppl. Table S2. Furthermore, the parity plots show that the DBAM predictions agree well with the hybrid MC/MD simulations for all elements (Suppl. Figure S11).

Interestingly, this DBAM provides a new physics insight via decoupling the effect of GB disorder on segregation. The second term $\Gamma_i^0(X) = \sum_i(\kappa_i^{Dis} \cdot X_i)$ in equation (3) is the composition contribution to the GB adsorption at the minimum disorder. The first term, $\beta_i \cdot (T - T_C) \cdot [\Gamma_{Dis}(T, X) - \Gamma_{Dis}^0]$, represents the "disorder contribution" (albeit it is in fact a coupled disorder and segregation effect). Thus, we can further quantify the fractions of this disorder contribution (the first term) to the 1032 model-predicted $\Gamma_i$ values and plot them in histograms for all five elements in Figure 5A. The large factions of 0.70 for $\Gamma_{Cr}$, 0.71 for $\Gamma_{Fe}$, and 0.66 for $\Gamma_{Co}$, respectively, suggest the significant roles of GB disorder in influencing the GB segregation of Cr, Fe, and Co. However, the fractions of disorder contributions are moderate (~0.46) for $\Gamma_{Mn}$ and almost zero for $\Gamma_{Ni}$ (Figure 4A). Interestingly, the fraction of disorder contribution is proportional to the absolute value of coupling coefficient $|\beta_i|$. These findings demonstrate the importance of GB disorder on influencing GB segregation in HEAs, particularly for Cr, Fe, and Co in this case.

The physical meaning and origin of $T_C$ are briefly summarized as follows. The equation (4) implies that at $T = T_C$, $\Gamma_i(T_C, X) = \Gamma_i^0(X)$, which is a small number (because $\Gamma_i^0 = \langle \Gamma_i^0(X) \rangle$ is small as shown in Figure 4C-F). The statistical analysis shown Figure 5A also suggests that $\Gamma_i^0(X)$ contributes only for small fractions to $\Gamma_i(T, X)$ for any strong segregating/depleting elements. Thus, $T_C$ represents the compensation temperature at which the effective GB segregation entropy ($\Delta S_{i \to 1}^{Seg\,(eff)}$, where the subscript "$i \to 1$" denotes the swap of an atom $i$ atom in the bulk and a reference atom 1 at the GB) is proportional to the effective GB segregation enthalpy ($\Delta H_{i \to 1}^{Seg(eff)}$) to produce $\Gamma_i(T_C, X) \sim 0$:

$$\Delta S_{i \to 1}^{Seg\,(eff)} = \frac{\Delta H_{i \to 1}^{Seg(eff)}}{T_C}. \tag{5}$$

Comparing with equation (4), we conclude that this entropic effect must be related to the increased GB disorder $\Delta \Gamma_{Dis}$. Thus, we can now envision the following picture for the physical meaning and origin of $T_C$. The increased GB disorder $\Delta \Gamma_{Dis}$ can reduce the effective GB free energy of segregation ($\Delta G_{i \to 1}^{Seg(eff)} = \Delta H_{i \to 1}^{Seg(eff)} - T \cdot \Delta S_{i \to 1}^{Seg(eff)}$) through the entropy of GB segregation, where the reduction is proportional to $\Delta H_{i \to 1}^{Seg(eff)}$ and more significant for strong segregating or depleting element. Thus, with increasing GB disorder $\Delta \Gamma_{Dis}$, GB segregation (or depletion) for different elements is reduced and equalized due to this entropic effect. The effective $\Delta G_{i \to 1}^{Seg(eff)}$ virtually vanishes (or is minimized) at this compensation temperature $T_C$. It should be noted that this compensation effect is likely only an approximated relation (because $\Gamma_i(X)$ is small but not exactly zero so that equation (5) is likely an approximation). Our data (Figure 4D and Suppl. Table S2) also show variations in the best fitted $T_C$ values for different elements ($\sim 1388 \pm 51$ K). We should also note that this predicted $T_C$ is from an extrapolation. As the temperature approaches the bulk solidus curve, premelting-like interfacial phases[6,29,33,34] can develop at GBs to change the projection. See Supplementary Discussion 6 for further elaboration about the origin and physical meaning of $T_C$.

It is interesting to further compare the fitted compositional coefficients ($\kappa_{i,j}^{Seg}$) with the corresponding segregation enthalpies in binary alloys. Also taking Cr as one example, Figure 5B shows the parity plot of Cr segregation enthalpies $\Delta H_{Cr,j}^{Seg}$ ($j = $ Mn, Fe, Co, Ni) *vs.* corresponding coupling coefficients $\kappa_{i,j}^{Seg}$, where an excellent linear relation with $R^2 = 0.95$ indicates a strong positive correlation. In addition, signs of the $\Delta H_{Cr,j}^{Seg}$ and $\kappa_{i,j}^{Seg}$ are always consistent. For instance, both a positive $\Delta H_{Cr,Fe}^{Seg}$ (or $\Delta H_{Cr,Co}^{Seg}$) in the classical segregation model and a positive $\kappa_{Cr,Fe}^{Seg}$ (or $\kappa_{Cr,Co}^{Seg}$) in our DBAM indicate preferred segregation of Cr at the GB of Fe (or Co). Likewise, negative $\Delta H_{Cr,Mn}^{Seg}$ (or $\Delta H_{Cr,Ni}^{Seg}$) and $\kappa_{Cr,Mn}^{Seg}$ (or $\kappa_{Cr,Ni}^{Seg}$) suggest preferred depletion of Cr at the GB of Mn (or Ni). Thus, the compositional coefficients ($\kappa_{i,j}^{Seg}$) are well correlated with binary segregation enthalpies.

Both the DBAM and ANN models can be used to map out the GB thermodynamic properties for HEAs in a 5D space as functions of four independent compositional DOFs and temperature. In comparison with the DBAM, the ANN model is more accurate for predicting GB properties with smaller RMSEs (Suppl. Tables S1-S2). But the ANN model predicts GB properties in a "black-box" approach without offering any physics insights; moreover, lacking the physical interpretation inhibits the model transferability. In contrast, the simple analytical formulae of the DBAM, where all model parameters have clear physical meanings, can provide understandings of the underlying physical interactions (including their signs and strengths in a quantitative way) in HEAs. Thus, this DBAM represents a general and transferrable GB thermodynamic model for HEAs, which can have significant and broad impacts.

**Comparisons with classical and other existing segregation models**

Here, we further compare the hybrid MC/MD-simulated GB segregation in HEAs (and the ANN and DBAM models derived based on this large MC/MD dataset) with segregation predicted by classical or other existing segregation models.

First, GB segregation in HEAs can exhibit more complex and intriguing behaviors than those in binary alloys. Here, we adopt the Wynblatt-Ku model[35] (considering both chemical and elastic contribution to GB segregation; see Supplementary Discussion 7) to compute GB fractions ($X_{GB}$) of Cr as functions of bulk fractions of Cr ($x = X_{Cr}$) for four $Cr_xM_{1-x}$ ($M = $ Mn, Fe, Co, Ni) binary alloys at 1000 K (Figure 5C). Then, we select four HEAs, including $Cr_xMn_{0.4-x}Fe_{0.2}Co_{0.2}Ni_{0.2}$ ($0.05 \leq x \leq 0.35$), as well as three variants where we swap Mn with Fe, Co, or Ni; we further plot MC/MD-simulated $\Gamma_{Cr}$ curves as functions of $x$ in Figure 5D. We notice several major differences in the trends of segregation in binary alloys vs. HEAs. The segregation strengths of Cr in binary alloys are ranked as Fe > Co > Ni > Mn (Figure 5C), while they are ranked as Mn ≈ Ni > Co ≈ Fe in HEAs (Figure 5D). More complex and intriguing compositional dependences, e.g., saturation of Cr segregation with $x > 0.2$ in $Cr_xMn_{0.2}Fe_{0.4-x}Co_{0.2}Ni_{0.2}$ and $Cr_xMn_{0.2}Fe_{0.2}Co_{0.4-x}Ni_{0.2}$ vs. acceleration of Cr segregation after $x > 0.2$ in $Cr_xMn_{0.2}Fe_{0.2}Co_{0.2}Ni_{0.4-x}$, are also observed in HEAs (Figure 5D).

Second, Li et al. recently proposed a density-based thermodynamic model for GB segregation in CrMnFeCoNi.[23] This phenomenological model assumed that GB energy can be written a function of GB density, which suggested the importance of GB free volume ($V_{Free}$). In contrast, the PCC heat map (Figure 4A) shows that $V_{Free}$ almost has no correlations with GB adsorption properties ($\Gamma_i$). Instead, $\Gamma_{Dis}$ exhibits strong correlation with $\Gamma_i$ (Figure 4A). Thus, we suggest that GB disorder (instead of density or free volume) should be treated as a key parameter for developing future phenomenological models. See Supplementary Discussion 8 for further discussion.

Third, we have extended a lattice-type model developed by Xing et al. for ternary alloys[36] to quinary HEAs. Although this model can predict some general trends, e.g., the positive segregation enthalpies for Cr, Mn, and Ni (segregation) vs. negative segregation enthalpies for Fe and Co (depletion), we cannot make quantitative predictions of GB segregation for non-equimolar HEAs. See Supplementary Discussion 9 for elaboration.

**Generality of the predictions**

In this study, most MC/MD simulations are based on an asymmetric Σ81 (mixed tilt and twist) GB to represent the behaviors of general GBs. To test generality of our predictions, we have also performed MC/MD simulations for three other GBs, including an asymmetric Σ15 (mixed tilt and twist) GB, a Σ41 symmetric tilt GB, and a Σ13 symmetric twist GB. For each of them, four non-equimolar HEAs selected based in the simulations of the asymmetric Σ81 GB diagrams, where the first three (HEA1-3) exhibit strong Cr segregation while last one (HEA4) has weak Cr segregation, were examined. Notably, MC/MD simulations show similar and consistent trends for all four GBs:

HEA1-3 always have large $\Gamma_{Cr}$, but HEA4 always has small $\Gamma_{Cr}$ (Suppl. Table S4 and Figure S15). Furthermore, DFT calculations also confirm that $E_{Seg}^{Cr}$ (around -0.026 eV/atom) of HEA1-3 is significantly lower than that for HEA4 (~ 0.0001 eV/atom), as shown in Supplementary Table S4. In conclusion, the trends predicted based on the asymmetric Σ81 (mixed tilt and twist) GB are likely representative. See Supplementary Discussion 10 for elaboration.

**Probing segregation mechanisms by first-principles calculations of electronic structures**

We have also calculated sum of bond ordering (SBO) values for the four non-equimolar HEAs discussed above to further understand how the bonding environment affects the Cr segregation (see Computational Procedures). Since SBO represents the total number of electrons that form bonds, similar SBO values indicate similar bonding environments. Interestingly, Fe, Cr, and Co atoms always have similar SBO values, which are ~4.04, ~3.95, and ~3.78, respectively. In contrast, Mn and Ni exhibit two distinct SBO values of ~4.20 and ~3.49, respectively (Suppl. Figure S16). Therefore, the preferred Cr segregation at the Fe- or Co-rich GBs can be understood because Fe or Co can provide more favorable segregation sites with similar bonding environments. On the other hand, the different bonding environments at Mn- or Ni-rich GBs can inhibit Cr segregation.

A recent study suggested that SBO can be used as a descriptor to predict and subsequently tailor GB segregation.[28] For example, if we want to promote segregation of a certain element (*e.g.*, Cr) in HEAs, we can increase the composition of the elements with similar SBO values (*e.g.*, Fe and Co) and/or reduce the composition of those with different SBO values (*e.g.*, Mn and Ni).

**Conclusions**

In this study, we used large-scale hybrid MC/MD simulations to generate a large dataset of GB properties for $Cr_xMn_yFe_zCo_lNi_m$ HEAs. The machine learning technique was firstly used to predict the GB properties of HEA as a function of four compositional DOFs and temperature in a 5D space. In addition, we found that interfacial disorder, as well as the interactions among segregation/depletion of five elements and GB disorder, can induce new and complex interfacial phenomena in HEAs, beyond the prediction of classical GB segregation theory. Notably, we discovered a GB critical compensation temperature in HEAs. Based on a careful analysis of the large dataset, we further created a DBAM to analytically represent GB segregation and disordering in the 5D space, where all parameters have clear physical meanings (*vs.* the black-box machine learning model). This work has enriched the classical GB segregation theory and developed a predictive and transferrable model for HEAs.

**Computational Procedures**

**Composition Selection and Principal Component Analysis (PCA).** In this work, the composition of each element was fixed in a range from 5 at% to 35 at% with a step of 5 at% for $Cr_xMn_yFe_zCo_lNi_m$. Since $x + y + z + l + m = 1$, there are 1371 possible compositions in total. Among them, we randomly selected 258 subsystems to perform high-throughput MC/MD simulations to generate a dataset. Principal component analysis (PCA) was used to analyze the composition distribution for these 258 subsystems to ensure the randomness of our selection, where the singular value decomposition (SVD) algorithm was chosen. The PCA were performed by Matlab2019a.

**Hybrid MC/MD simulation for GB diagrams.** The GBstudio[37] website was used to construct a mixed twist-tilt Σ81 GB with boundary planes $(1\bar{1}0)//(7\bar{8}7)$ to represent general GBs. The energy minimization for each GB was first performed at 0 K by conjugate gradient (CG) algorithm. Subsequently, the hybrid Monte Carlo and molecular dynamics (hybrid MC/MD) simulations in constant *NPT* ensembles were carried out to swap atoms and find energetically favorable GB structure. Five MC trial moves were conducted between each MD step with a 0.1 fs MD time step and $10^5$ hybrid MC/MD steps performed for each simulation to achieve convergence. All MC/MD simulations were performed using the LAMMPS code.[38] A 2NN MEAM potential[39] was adopted for CrMnFeCoNi alloys.

The methods used to calculate GB excess of solute (*i.e.*, GB adsorption amount $\Gamma_{Cr}$, $\Gamma_{Mn}$, $\Gamma_{Fe}$, $\Gamma_{Co}$, $\Gamma_{Ni}$) and disorder ($\Gamma_{Dis}$) diagrams were same as our prior studies; the detailed procedures are described in Refs. 20,28. To calculate GB free volume ($V_{Free}$), we used the relation of $V_{Free} = V_{Total} \cdot \sum \Gamma_i$, where $V_{Total}$ is the total volume of GB structure and $i$ = Cr, Mn, Fe, Co, or Ni. To minimize the thermal noise effect, we calculated each GB property based on the average of five random structures during the last five MC/MD steps.

It should be noted that we set an overall global composition in a hybrid MC/MD simulation. The bulk composition is recalculated based the grain composition (away from the GB region) after achieving the chemical equilibrium, which is subsequently used for both training the ANN model and further analysis and developing a data-based analytical model.

**Artificial neural networks.** The data set was divided into training, validation, and test subsets in a ratio of 0.7:0.15:0.15. The Levenberg-Marquardt backpropagation function was adopted to train ANN models. We found the optimized network architectures for the ANN ($n^i$-$n^{[i]}$-1, where $n^i$ is the number of input parameter, $n^{[i]}$ (the number of neurons in the single layer) is set to be 6-20-1. All data processing and ANN development were performed by Deep Learning Toolbox in Matlab2019a.

**Derivation of the Data-Based Analytical Model (DBAM).** Based on the linear regression analyses shown in Figure 4C and Suppl. Figure S10, the adsorption amount, $\Gamma_i(T, X)$, is statistically correlated with the GB excess of disorder, $\Gamma_{Dis}(T, X)$, linearly with the slope $\bar{\alpha}^i_{Dis}(T)$ at a given temperature $T$, where $X = \{X_i\}$ is a concise form to note the bulk composition of the HEA. Thus, we statistically have the following linear correlation:

$$\Gamma_i(T, X) - \Gamma_i^0 = \bar{\alpha}^i_{Dis}(T) \cdot [\Gamma_{Dis}(T, X) - \Gamma_{Dis}^0], \tag{6}$$

where ($\Gamma_i^0$, $\Gamma_{Dis}^0$) is the intersection point of all linear regression lines for different temperatures in each panel of Figure 4C, and they are virtually independent of temperature. We can observe in Figure 4C that $\Gamma_i^0$ is a relatively small number: $\Gamma_i^0 \ll \langle \Gamma_i(T,X) \rangle$. Furthermore, the linear regression analyses shown in Figure 4D suggest:

$$\bar{\alpha}_{Dis}^i(T) = \beta_i \cdot (T - T_C), \tag{7}$$

where $\beta_i$ is slope of the linear regression line in Figure 4D. Here, $T_C \approx 1388 \pm 51$ K is a critical temperature shown in Figure 4D. At $T = T_C$, $\Gamma_i(T_C, X) = \Gamma_i^0 \sim 0$ (see Suppl. Figure S10), so this critical temperature is the so-called "compensation" temperature of GB segregation.[40] Equations (6)-(7) are same as equations (1)-(2) in the main text. We further propose:

$$\Gamma_i(T,X) = \beta_i \cdot (T - T_C) \cdot [\Gamma_{Dis}(T,X) - \Gamma_{Dis}^0] + \Gamma_i^0(X). \tag{8}$$

Here, we can assume $\Gamma_{Dis}^0 = \Gamma_{Dis}^{min}$ (or the minimum among all possible HEAs compositions), which are approximately held based on Figure 4C and Suppl. Figure S10, except for the case of Ni, where there are too high noises due to the small values of $\Gamma_{Ni}$. In equation (6) and Figure 4 and S10, $\Gamma_i^0$ is a fitted constant independent of $X$. In equation (9), we further generalize equation (6) to allow this constant $\Gamma_i^0$ to be a function of $X$ to enable more accurate fitting, where we have $\langle \Gamma_i^0(X) \rangle \sim \Gamma_i^0$. Here, we may adopt a linear expression as a first-order approximation:

$$\Gamma_i^0(X) = \sum_j \left( \kappa_{i,j}^{Seg} \cdot X_j \right) \tag{9}$$

where $\kappa_{i,j}^{Seg}$ is a coupling coefficient for the GB segregation. Thus, we have:

$$\Gamma_i(T,X) = \beta_i \cdot (T - T_C) \cdot [\Gamma_{Dis}(T,X) - \Gamma_{Dis}^0] + \sum_j \left( \kappa_{i,j}^{Seg} \cdot X_j \right). \tag{10}$$

Since GB disorder should increase with temperature, we propose the following relation:

$$\Gamma_{Dis}(T,X) = \Gamma_{Dis,0}(X) \cdot \exp\left(-\frac{E_A^{Dis}}{k_B T}\right), \tag{11}$$

where $E_A^{Dis}$ is the activation energy of disordering, and $k_B$ is the Boltzmann constant. We again adopt a linear expression as a first-order approximation for the temperature-independent pre-factor:

$$\Gamma_{Dis,0}(X) = \sum_i \left( \kappa_i^{Dis} \cdot X_i \right) \tag{12}$$

Next, we can use all hybrid MC/MD-simulated data points to fit equations (11) and (12). Finally, by combining equations (10)-(12), we can obtain:

$$\Gamma_i(T,X) = \beta_i \cdot (T - T_C) \cdot \left[ \sum_i (\kappa_i^{Dis} \cdot X_i) \exp\left(-\frac{E_A^{Dis}}{k_B T}\right) - \Gamma_{Dis}^0 \right] + \sum_j \left( \kappa_{i,j}^{Seg} \cdot X_j \right). \tag{13}$$

Further discussions about the DBAM model and the physical meaning and origin of $T_C$ can be found in Supplementary Discussions 5 and 6.

**Density function theory (DFT) calculations.** The first-principles DFT calculations were performed by using the Vienna *ab initio* Simulations Package (VASP).[41,42] The Kohn-Sham equations were used to solve the projected-augmented wave (PAW) method[43,44] along with standard PAW potentials. All GB structures were fully relaxed until the Hellmann-Feynman forces were smaller than 0.02 eV/Å. The Brillouin-zone integrations were sampled on a $\Gamma$-centered 2×2×1 $k$-point grids. The kinetic energy cutoff for plane waves was set to 368 eV. The convergence criterion for the electronic self-consistency was set to $10^{-4}$ eV. The "high" precision setting was adopted to avoid wrap around errors. The spin-polarization was not considered due to weak effect on atomic arrangement.[45] The SBO was calculated by using the state-of-the-art DDEC06 method[46] following the all-electron static calculations.

**Acknowledgements:** This work is currently supported by the UC Irvine MRSEC, Center for Complex and Active Materials, under National Science Foundation award DMR-2011967. We also acknowledge the earlier support from a Vannevar Bush Faculty Fellowship sponsored by the Basic Research Office of the Assistant Secretary of Defense for Research and Engineering and funded through the Office of Naval Research under Grant No. N00014-16-2569, prior to the establishment of the MRSEC. The calculations were performed at the Triton Shared Computing Cluster (TSCC) at the University of California, San Diego (UCSD).

**Author contributions:** J. L. conceived the idea and supervised the work. C. H. performed simulations and calculations. Both authors wrote, reviewed, and revised the manuscript.

**Competing interests:** The authors declare no conflict of interests.

**Data availability:** The MC/MD-simulated dataset for DBAM and ANN models are available on GitHub website https://github.com/huhuhhhh/HEAGB/tree/main/raw_data. The supporting data for generating binary and ternary GB diagrams are also available in this repository.

**Code availability:** The MATLAB scripts for fitting DBAM and training ANN models are available on GitHub website https://github.com/huhuhhhh/HEAGB/tree/main.

**Supplementary Information:**
Supplementary Discussions 1-12
Supplementary Tables S1-S4
Supplementary Figures S1–S29

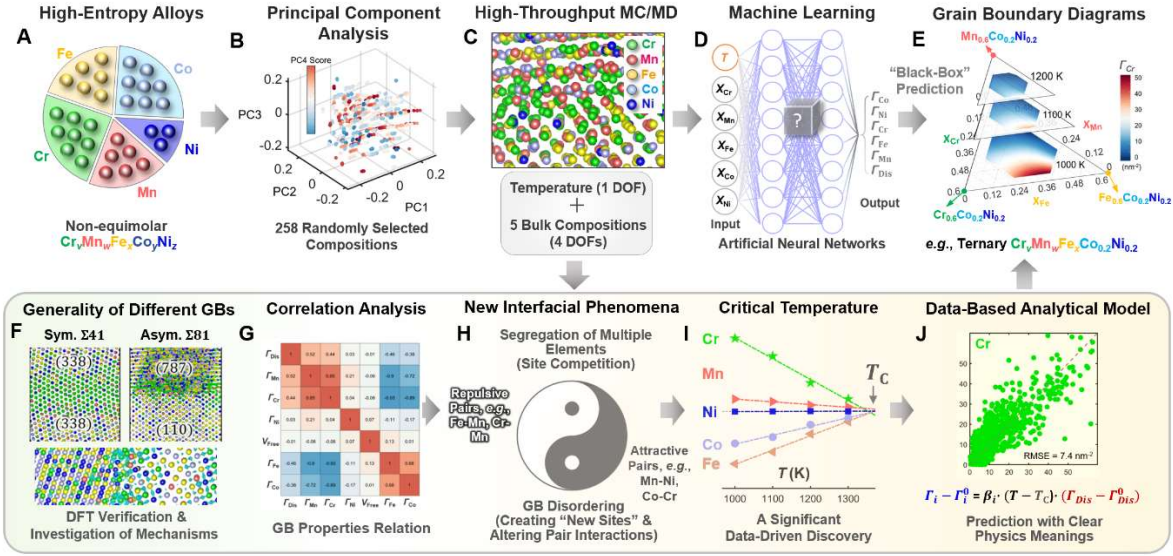

**Figure 1. Workflow of the machine learning prediction of grain boundary (GB) diagrams, data-based discovery of new interfacial phenomena, and development of a data-based analytical model (DBAM).**
(A) Schematic diagram of non-equimolar five-element $Cr_xMn_yFe_zCo_lNi_m$ alloys.
(B) Principal component analysis (PCA) verifying the randomness in the selection of 258 HEAs.
(C) The equilibrium structure of an asymmetric Σ81 GB in $Co_{0.2}Ni_{0.2}Cr_{0.2}Fe_{0.35}Mn_{0.05}$ at 1000 K obtained by hybrid Monte Carlo and molecular dynamics (hybrid MC/MD) simulations. In total, 1032 such individual hybrid MC/MD simulations were performed for 258 HEAs at four different temperatures to calculate GB excesses of solutes (*i.e.*, $\Gamma_{Cr}$, $\Gamma_{Mn}$, $\Gamma_{Fe}$, $\Gamma_{Co}$, $\Gamma_{Ni}$) and disorder ($\Gamma_{Dis}$), and free volume ($V_{Free}$).
(D) Schematic diagram of an artificial neural networks (ANN) for predicting six GB properties.
(E) An example of GB diagrams predicted by the ANN model for a ternary $Cr_xMn_yFe_zCo_{0.2}Ni_{0.2}$ ($x + y + z = 0.6$) subsystem, showing three isothermal sections of the Cr adsorption ($\Gamma_{Cr}$) diagrams.
(F) Screenshot of strong Cr segregation in different GBs, which is also verified by DFT calculations.
(G) Correlation analysis of GB properties.
(H) Schematic of new interfacial phenomena in HEAs.
(I) A data-driven discovery of a GB critical temperature $T_C$, leading to the development of a DBAM.
(J) The surrogate DBAM that can predict GB properties with parameters of clear physical meanings.

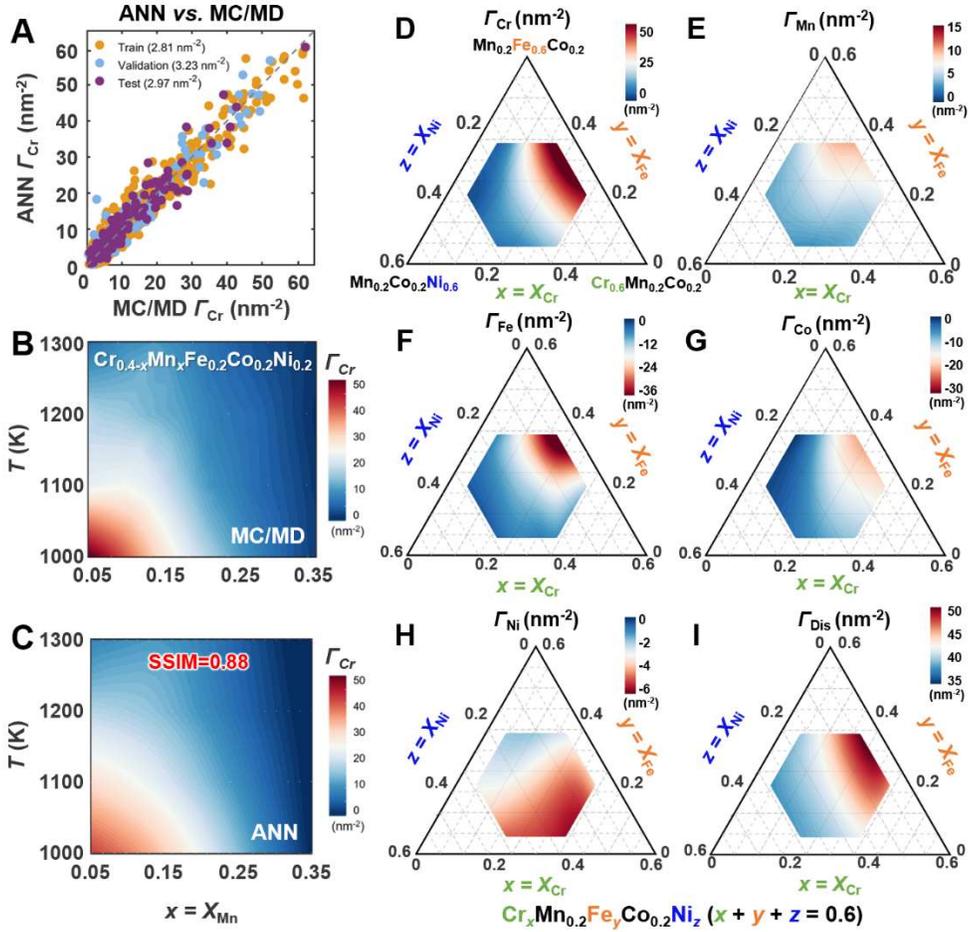

**Figure 2. ANN performance for predicting GB properties.**
(A) Parity plot of ANN predictions *vs.* MC/MD simulations for the GB excess of Cr adsorption ($\Gamma_{Cr}$).
(B-C) MC/MD-simulated *vs.* ANN-predicted isopleths of $\Gamma_{Cr}$ diagrams as functions of temperature and Mn bulk composition ($x = X_{Mn}$) for the $Cr_{0.4-x}Mn_xFe_{0.2}Co_{0.2}Ni_{0.2}$ system.
(D-I) Representative ternary isothermal sections of ANN-predicted GB diagrams of $\Gamma_{Cr}$, $\Gamma_{Mn}$, $\Gamma_{Fe}$, $\Gamma_{Co}$, $\Gamma_{Ni}$, and $\Gamma_{Dis}$ for $Cr_xMn_{0.2}Fe_yCo_{0.2}Ni_z$ ($x + y + z = 0.6$; $x = X_{Cr}$, $y = X_{Fe}$, $z = X_{Ni}$) at 1000 K.

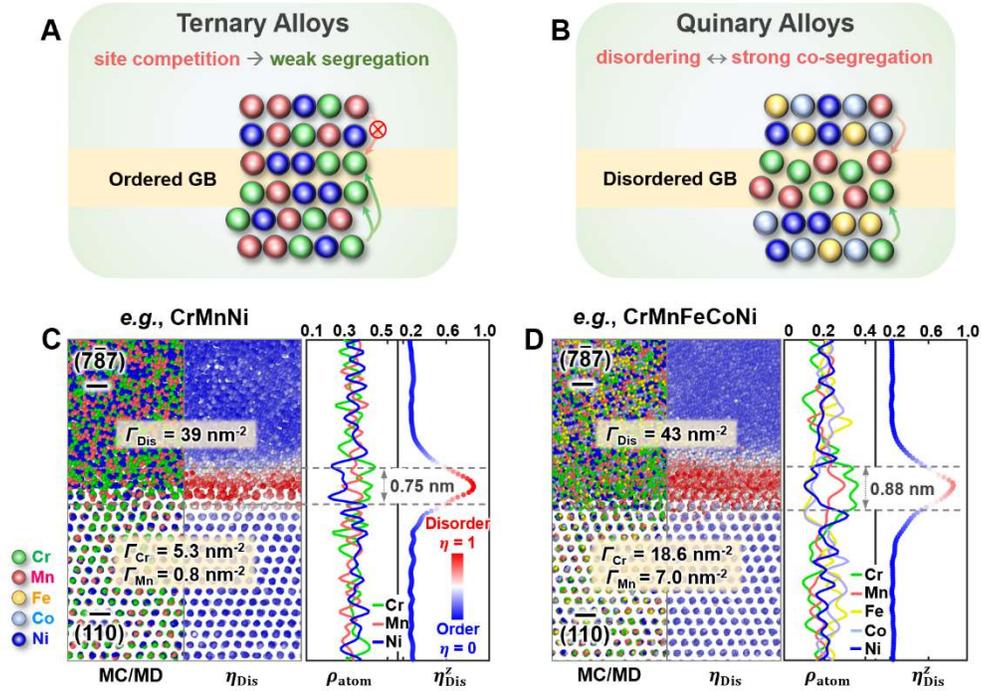

**Figure 3. Unique coupled interfacial disordering and GB co-segregation behaviors in HEAs, exemplified by comparing the same asymmetric Σ81 GB in equimolar CrMnNi *vs.* CrMnFeCoNi alloys at 1000 K.**

(A) Schematic of weak segregation in ternary alloys ascribed to the site competition in relatively ordered GBs.

(B) Schematic of the coupling of interfacial disordering and strong co-segregation of multiple elements in quinary alloys.

(C) MC/MD-simulated GB structure of the equimolar CrMnNi alloy and the corresponding disorder parameter ($\eta_{Dis}$) and atomic density profiles. This GB exhibits an GB excess disorder $\Gamma_{Dis}$ of ~39 nm$^{-2}$, moderate segregation of Cr ($\Gamma_{Cr}$ = ~5.3 nm$^{-2}$), and weak segregation of Mn ($\Gamma_{Mn}$ = ~0.8 nm$^{-2}$).

(D) MC/MD-simulated GB structure of the equimolar CrMnFeCoNi and the corresponding disorder parameter ($\eta_{Dis}$) and atomic density profiles. In comparison with the same GB in the ternary CrMnCr alloy, this GB in the quinary Cantor alloy is more disordered with a larger $\Gamma_{Dis}$ of ~43 nm$^{-2}$ and strong co-segregation of Cr and Mn ($\Gamma_{Cr}$ = ~18.3 nm$^{-2}$ and $\Gamma_{Cr}$ = ~7.0 nm$^{-2}$, which represent ~3.5× and ~9× increases, respectively, from those in the ternary alloy).

More examples and further discussion can be found in Supplementary Discussion 3.

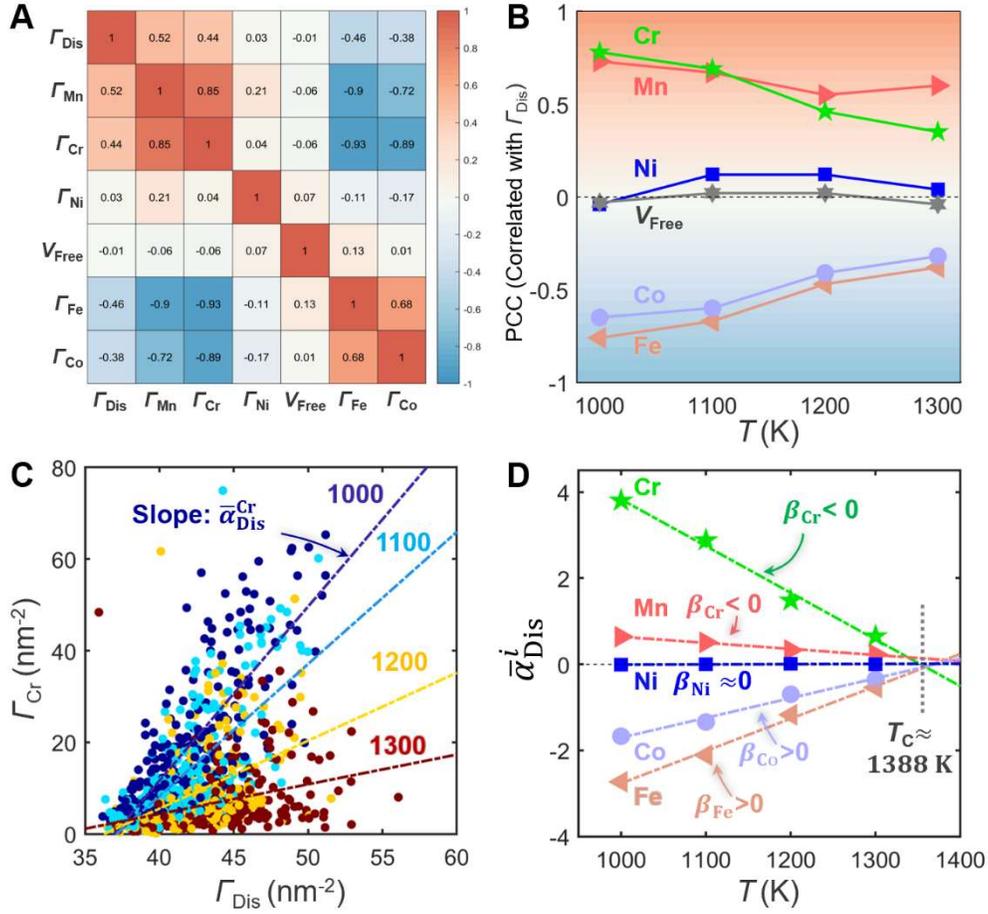

**Figure 4. Correlation analyses of GB thermodynamic properties.**

(A) Heat map of Pearson correlation coefficients (PCCs) between all pairs of the seven GB properties.

(B) Calculated correlation coefficients between GB excess of disorder ($\varGamma_{Dis}$) and six other GB properties (GB excesses of Cr, Fe, Co, Ni and Mn, as well as GB free volume) at different temperatures.

(C) Plots of GB excess of Cr ($\varGamma_{Cr}$) vs. GB excess of disorder ($\varGamma_{Dis}$) at 1000 K, 1100 K, 1200 K, and 1300 K, respectively, for $i$ = (c) Cr, (d) Mn, (e) Fe, and (f) Co. The dashed lines are regression lines of $\varGamma_{Cr}$ vs. $\varGamma_{Dis}$ at different temperatures based on 258 HEA compositions. The plot for other four adsorption properties vs. $\varGamma_{Dis}$ can be found in Figure S10. The slopes of these dashed lines are labelled as $\bar{\alpha}_{Dis}^{i}$, where $i$ = Cr, Mn, Fe, Co, and Ni.

(D) The fitted $\bar{\alpha}_{Dis}^{i}$ as functions of temperature ($T$) for five elements. The slopes of $\bar{\alpha}_{Dis}^{i}$ vs. $T$ regression lines are labelled as $\beta_i$. Notably, all five fitted linear lines cross over at nearly one point on the horizontal $T$-axis ($\bar{\alpha}_{Dis}^{i} = 0$) at $T_C \approx 1388$ K.

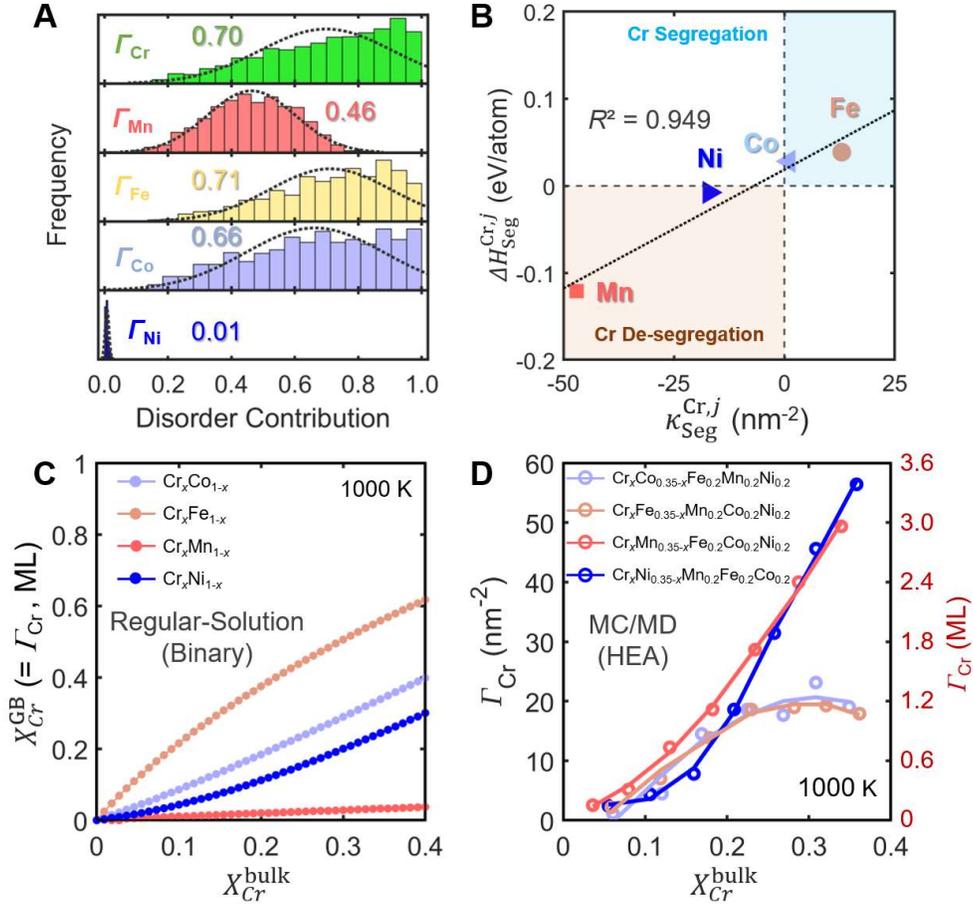

**Figure 5**. **Analyses and comparisons of the data-based analytical model (DBAM), classical segregation models, and MC/MD simulations**.

(A) Histograms of the disorder contribution to GB adsorption of each element based on the DBAM.

(B) Parity plot of $\Delta H_{Cr,j}^{Seg}$ (the segregation enthalpy of Cr in the binary Cr-$j$ alloy, where $j$ = Mn, Fe, Co, and Ni, calculated used a lattice-type model [47]) vs. $\kappa_{Seg}^{Cr,j}$ (the compo_ ENREF_6sitional coupling coefficients in the DBAM). The positive (or negative) values of $\Delta H_{Cr,j}^{Seg}$ or $\kappa_{Cr,j}^{Seg}$ indicate Cr is favorable (or unfavorable) to segregate.

(C) Calculated GB adsorption ($X_{Cr}^{GB} = \Gamma_{Cr}$) vs. the bulk Cr fraction ($X_{Cr}^{bulk}$) for four $Cr_x$-$j_{1-x}$ ($j$ = Mn, Fe, Co, and Ni) binary alloys at 1000 K using the Wynblatt-Ku model [35].

(D) MC/MD-simulated $\Gamma_{Cr}$ vs. $X_{Cr}^{bulk}$ for four HEAs at 1000 K. The compositions are noted in the legend, where in each case the increase in the Cr fraction is compensated by one selected element while keeping the fractions of the three other elements at the constant level of 0.2.

# References

1  Ye, Y.F., Wang, Q., Lu, J., Liu, C.T., and Yang, Y. (2016). High-entropy alloy: Challenges and prospects. *Materials Today* **19**, 349-362. https://doi.org/10.1016/j.mattod.2015.11.026
2  Tsai, M.-H., and Yeh, J.-W. (2014). High-entropy alloys: A critical review. *Materials Research Letters* **2**, 107-123. 10.1080/21663831.2014.912690
3  Miracle, D.B., and Senkov, O.N. (2017). A critical review of high entropy alloys and related concepts. *Acta Materialia* **122**, 448-511. 10.1016/j.actamat.2016.08.081
4  George, E.P., Raabe, D., and Ritchie, R.O. (2019). High-entropy alloys. *Nature Reviews Materials* **4**, 515-534. 10.1038/s41578-019-0121-4
5  Dillon, S.J., Tang, M., Carter, W.C., and Harmer, M.P. (2007). Complexion: A new concept for kinetic engineering in materials science. *Acta Materialia* **55**, 6208-6218. https://doi.org/10.1016/j.actamat.2007.07.029
6  Cantwell, P.R., Tang, M., Dillon, S.J., Luo, J., Rohrer, G.S., and Harmer, M.P. (2014). Grain boundary complexions. *Acta Materialia* **62**, 1-48. https://doi.org/10.1016/j.actamat.2013.07.037
7  Dey, D., and Bradt, R.C. (1992). Grain growth of zno during bi2o3 liquid-phase sintering. *Journal of the American Ceramic Society* **75**, 2529-2534. 10.1111/j.1151-2916.1992.tb05607.x
8  Nie, J., Chan, J.M., Qin, M., Zhou, N., and Luo, J. (2017). Liquid-like grain boundary complexion and sub-eutectic activated sintering in cuo-doped tio2. *Acta Materialia* **130**, 329-338. https://doi.org/10.1016/j.actamat.2017.03.037
9  Luo, J., Wang, H., and Chiang, Y.-M. (1999). Origin of solid-state activated sintering in bi2o3-doped zno. *Journal of the American Ceramic Society* **82**, 916-920. 10.1111/j.1151-2916.1999.tb01853.x
10  Dillon, S.J., Tai, K., and Chen, S. (2016). The importance of grain boundary complexions in affecting physical properties of polycrystals. *Current Opinion in Solid State and Materials Science* **20**, 324-335. https://doi.org/10.1016/j.cossms.2016.06.003
11  Krause, A.R., Cantwell, P.R., Marvel, C.J., Compson, C., Rickman, J.M., and Harmer, M.P. (2019). Review of grain boundary complexion engineering: Know your boundaries. *Journal of the American Ceramic Society* **102**, 778-800. 10.1111/jace.16045
12  Hu, T., Yang, S., Zhou, N., Zhang, Y., and Luo, J. (2018). Role of disordered bipolar complexions on the sulfur embrittlement of nickel general grain boundaries. *Nature communications* **9**, 2764
13  Westbrook, J.H. (1964). Segregation at grain boundaries. *Metallurgical Reviews* **9**, 415-471. 10.1179/mtlr.1964.9.1.415
14  McLean, D. (1957). Grain boundaries in metals. *Oxford, Clarendon Press*
15  Hondros, E.D., and Seah, M.P. (1977). The theory of grain boundary segregation in terms of surface adsorption analogues. *Metallurgical Transactions A* **8**, 1363-1371. 10.1007/BF02642850
16  Pan, Z., and Rupert, T.J. (2016). Effect of grain boundary character on segregation-induced structural transitions. *Physical Review B* **93**, 134113. 10.1103/PhysRevB.93.134113
17  Frolov, T., Divinski, S.V., Asta, M., and Mishin, Y. (2013). Effect of interface phase transformations on diffusion and segregation in high-angle grain boundaries. *Physical Review Letters* **110**, 255502. 10.1103/PhysRevLett.110.255502
18  Frolov, T., Asta, M., and Mishin, Y. (2015). Segregation-induced phase transformations in grain boundaries. *Physical Review B* **92**, 020103. 10.1103/PhysRevB.92.020103
19  Hu, C., and Luo, J. (2019). First-order grain boundary transformations in au-doped si: Hybrid monte carlo and molecular dynamics simulations verified by first-principles calculations. *Scripta Mater.* **158**, 11-15. https://doi.org/10.1016/j.scriptamat.2018.08.017
20  Yang, S., Zhou, N., Zheng, H., Ong, S.P., and Luo, J. (2018). First-order interfacial transformations with a critical point: Breaking the symmetry at a symmetric tilt grain boundary. *Phys. Rev. Lett.* **120**, 085702
21  Ming, K., Li, L., Li, Z., Bi, X., and Wang, J. (2019). Grain boundary decohesion by nanoclustering ni and cr separately in crmnfeconi high-entropy alloys. *Science Advances* **5**, eaay0639. 10.1126/sciadv.aay0639
22  Li, Y.J., Savan, A., Kostka, A., Stein, H.S., and Ludwig, A. (2018). Accelerated atomic-scale exploration of phase evolution in compositionally complex materials. *Materials Horizons* **5**, 86-92. 10.1039/C7MH00486A
23  Li, L., Kamachali, R.D., Li, Z., and Zhang, Z. (2020). Grain boundary energy effect on grain boundary segregation in an equiatomic high-entropy alloy. *Physical Review Materials* **4**, 053603. 10.1103/PhysRevMaterials.4.053603
24  Wynblatt, P., and Chatain, D. (2019). Modeling grain boundary and surface segregation in multicomponent high-entropy alloys. *Physical Review Materials* **3**, 054004. 10.1103/PhysRevMaterials.3.054004

25 Lee, H., Shabani, M., Pataky, G.J., and Abdeljawad, F. (2021). Tensile deformation behavior of twist grain boundaries in cocrfemnni high entropy alloy bicrystals. *Scientific reports* **11**

26 Tang, M., Carter, W.C., and Cannon, R.M. (2006). Grain boundary transitions in binary alloys. *Physical Review Letters* **97**, 075502. 10.1103/PhysRevLett.97.075502

27 Hart, E.W. (1968). 2-dimensional phase transformation in grain boundaries. *Scripta Metall.* **2**, 179-&. 10.1016/0036-9748(68)90222-6

28 Hu, C., Zuo, Y., Chen, C., Ping Ong, S., and Luo, J. (2020). Genetic algorithm-guided deep learning of grain boundary diagrams: Addressing the challenge of five degrees of freedom. *Materials Today*. https://doi.org/10.1016/j.mattod.2020.03.004

29 Zhou, N., Hu, T., and Luo, J. (2016). Grain boundary complexions in multicomponent alloys: Challenges and opportunities. *Current Opinion in Solid State and Materials Science* **20**, 268-277. https://doi.org/10.1016/j.cossms.2016.05.001

30 Morawiec, A., and Glowinski, K. (2013). On "macroscopic" characterization of mixed grain boundaries. *Acta Materialia* **61**, 5756-5767. https://doi.org/10.1016/j.actamat.2013.06.019

31 Li, L., Li, Z., Kwiatkowski da Silva, A., Peng, Z., Zhao, H., Gault, B., and Raabe, D. (2019). Segregation-driven grain boundary spinodal decomposition as a pathway for phase nucleation in a high-entropy alloy. *Acta Materialia* **178**, 1-9. https://doi.org/10.1016/j.actamat.2019.07.052

32 Larsen, P.M., Schmidt, S., and Schiøtz, J. (2016). Robust structural identification via polyhedral template matching. *Modelling and Simulation in Materials Science and Engineering* **24**, 055007. 10.1088/0965-0393/24/5/055007

33 Luo, J., and Shi, X.M. (2008 ). Grain boundary disordering in binary alloys. *Applied Physics Letters* **92**, 101901 10.1063/1.2892631

34 Luo, J. (2012). Developing interfacial phase diagrams for applications in activated sintering and beyond: Current status and future directions. *J. Am. Ceram. Soc.* **95**, 2358-2371

35 Wynblatt, P., and Ku, R.C. (1977). Surface energy and solute strain energy effects in surface segregation. *Surface Science* **65**, 511-531. https://doi.org/10.1016/0039-6028(77)90462-9

36 Xing, W., Kalidindi, A.R., Amram, D., and Schuh, C.A. (2018). Solute interaction effects on grain boundary segregation in ternary alloys. *Acta Materialia* **161**, 285-294. https://doi.org/10.1016/j.actamat.2018.09.005

37 Ogawa, H. (2006). Gbstudio: A builder software on periodic models of csl boundaries for molecular simulation. *MATERIALS TRANSACTIONS* **47**, 2706-2710. 10.2320/matertrans.47.2706

38 Plimpton, S. (1995). Fast parallel algorithms for short-range molecular dynamics. *Journal of Computational Physics* **117**, 1-19. https://doi.org/10.1006/jcph.1995.1039

39 Choi, W.-M., Jo, Y.H., Sohn, S.S., Lee, S., and Lee, B.-J. (2018). Understanding the physical metallurgy of the cocrfemnni high-entropy alloy: An atomistic simulation study. *npj Computational Materials* **4**, 1. 10.1038/s41524-017-0060-9

40 Wynblatt, P., and Chatain, D. (2006). Anisotropy of segregation at grain boundaries and surfaces. *Metallurgical and Materials Transactions a-Physical Metallurgy and Materials Science* **37A**, 2595-2620. 10.1007/bf02586096

41 Kresse, G., and Hafner, J. (1993). Ab initio molecular dynamics for liquid metals. *Phys. Rev. B* **47**

42 Kresse, G., and Furthmüller, G. (1996). Efficient iterative schemes for ab initio total-energy calculations using a plane-wave basis set. *Phys. Rev. B* **54**, 11169-11186

43 Blöchl, P.E. (1994). Projector augmented-wave method. *Phys. Rev. B* **50**, 17953-17979

44 Kresse, G., and Joubert, D. (1999). From ultrasoft pseudopotentials to the projector augmented-wave method. *Phys. Rev. B* **59**

45 Leong, Z., Wróbel, J.S., Dudarev, S.L., Goodall, R., Todd, I., and Nguyen-Manh, D. (2017). The effect of electronic structure on the phases present in high entropy alloys. *Scientific Reports* **7**, 39803. 10.1038/srep39803

46 Manz, T.A. (2017). Introducing ddec6 atomic population analysis: Part 3. Comprehensive method to compute bond orders. *RSC Advances* **7**, 45552-45581. 10.1039/C7RA07400J

47 Murdoch, H.A., and Schuh, C.A. (2013). Estimation of grain boundary segregation enthalpy and its role in stable nanocrystalline alloy design. *Journal of Materials Research* **28**, 2154-2163. 10.1557/jmr.2013.211